\documentclass[12pt]{article}
\usepackage{geometry}
\usepackage{amsmath, amssymb}
\usepackage{graphicx}
\usepackage{booktabs}
\usepackage{multirow}
\usepackage{array}
\usepackage{longtable}
\usepackage{caption}
\usepackage{setspace}
\usepackage{float}
\usepackage{indentfirst}
\usepackage{natbib}
\usepackage{hyperref}
\usepackage{titlesec}
\usepackage{multirow}
\usepackage{booktabs}
\usepackage{array}
\usepackage{threeparttable}
\usepackage{tabularx}
\usepackage{makecell}
\usepackage{booktabs}
\usepackage{amsmath}
\usepackage{graphicx}
\usepackage{array}
\newcolumntype{Y}{>{\centering\arraybackslash}X}
\usepackage{natbib}
\titleformat{\section}{\large\bfseries}{\thesection}{1em}{}
\titleformat{\subsection}{\normalsize\bfseries}{\thesubsection}{1em}{}
\titleformat{\subsubsection}{\normalsize\bfseries}{\thesubsubsection}{1em}{}

\title{An Investigation into the Applicability of Friction Velocity Estimation Methods for the Channel with A Deposit Body}

\author{
\begin{tabular}{l}
Jing Zhang\textsuperscript{1,3,*}, Ruihan Qin\textsuperscript{1,3}, Zhixue Guo\textsuperscript{2}, Huantao Xu\textsuperscript{1,3}, Jiahui Zhou\textsuperscript{1,3}\\Yun Bai\textsuperscript{1,3} \\[1pt]
\small\textsuperscript{1}School of Energy and Power Engineering, Key Laboratory of Fluid and Power Machinery, \\
\small\hspace{0.5cm}Ministry of Education, Xihua University, Chengdu, 610039, China \\[1pt]
\small\textsuperscript{2}State Key Laboratory of Hydraulics and Mountain River Engineering, \\
\small\hspace{0.5cm}Sichuan University, Chengdu, 610065, China \\[1pt]
\small\textsuperscript{3}School of Energy and Power Engineering, Xihua University, Chengdu, 610039, China \\[1pt]
\small{*Corresponding author: zhangjing\_w0521@vip.163.com}
\end{tabular}
}

\begin{document}
\maketitle
\begin{abstract}
This study investigates how the deposit body influences friction characteristics by altering local flow fields, which is closely related to bed shear stress. Using generalized flume experiments, the study assesses the applicability of classical uniform flow friction models in deposit body river sections, revealing frictional changes induced by flow field non-uniformity. Initially, based on \( \frac{u_*}{\sqrt{\overline{w'^2}}} = 0.85 \sim 1.15 \) under uniform flow conditions as the judgment basis, the reliability of the classical model is verified. Four models are then applied to estimate near-bed friction velocity in deposit body sections. Results show a significant alignment between the longitudinal velocity gradient and the peak friction velocity derived from the turbulent kinetic energy method (TKE). Dimensional analysis of friction indicators reveals that: (a) friction velocity is primarily influenced by turbulence intensity, with constricted and narrowed sections resembling uniform flow, while the expansion section forms a peak; (b) models incorporating flow field fluctuations (TKE, Vertical Turbulence Kinetic Energy (TKE w'), Reynolds shear stress method (RSS)) effectively capture the impact of non-uniform flow fields on friction characteristics; (c) when energy states are low or when deposit body proportions are large, the deposit body's resistance ratio increases, and peak friction velocity rises. This study provides theoretical insights into friction estimation and sediment transport in non-uniform flow fields of deposit bodies.

\end{abstract}

\section{Introduction}
Approximately 200,000 landslides were triggered by the May 12, 2008 Wenchuan earthquake and subsequent strong rainfall events \cite{xu2014three}. After an earthquake in mountainous areas, slope stability is reduced, leading to large amounts of landslide deposits accumulating in rivers or on slopes. When these loose deposits are transported to rivers, they may form bank deposits, even blocking the river and creating landslide dams, triggering a series of geological disasters. Wenchuan is located in a subtropical humid climate zone, influenced by the southwest and southeast monsoons in summer. The climate is relatively humid, with rainfall concentrated from July to September, characterized by frequent rain days and concentrated rainfall time and areas. Multiyear imagery of the Wenchuan earthquake zone from 2008 to 2019 reveals that although the volume of landslide-derived sediment supplied to river channels has significantly decreased in recent years, large quantities of landslide sediment continue to be transported into the river during periods of heavy rainfall. Moreover, substantial amounts of sediment remain in both the tributary and main river channels \cite{xiong2022long}.The deposit body significantly alter river flow by constricting space, affecting sediment transport and riverbed evolution, which in turn threatens flood discharge and navigation safety. While previous studies have explored changes in flow, water surface, and flow patterns of the deposit body \cite{he2017influence, huang2017experimental, zhou2018experimental}, few have addressed bedload movement about riverbed evolution. Accurately measuring bed shear stress in non-uniform flow under accumulation influence is key to solving this problem. The precision of direct bed shear stress measurements is heavily reliant on the methods, tools, and techniques employed \cite{aliaga2024bed, park2016direct, pujara2014direct}. Bed shear stress is usually indirectly estimated using friction velocity ($\tau =\rho {{u}_{*}}$). Over the past decade, researchers have focused on frictional flow velocity estimation models. By introducing parameters reflecting flow conditions or local bed micro-mechanisms, they have developed empirical formulas or new models \cite{li2024semi, rehman2022influence, chen2010sediment}, expanding the formulas' applicability to various flow conditions.As computational power improves, numerical simulations have been applied to open channel flow. Using models such as RANS and LES, Mishra and Venayagamoorthy (2024) and Etminan (2018) analyzed frictional flow velocity under complex conditions to improve sediment transport predictions \cite{mishra2024new, etminan2018predicting}. Current research on resistance in complex flows still largely relies on uniform flow models.

Table 1 summarizes six classic methods for calculating uniform flow frictional velocity \cite{graf1995bed, soulsby1983chapter, kim2000estimating, schlichting1980boundary, wang2001velocity}. Bagherimiyab and Lemmin (2013) performed experiments on a coarse gravel bed (D\textsubscript{50} = 1.5 cm) and found that the wall similarity method differed by less than 10\% from the TKE and RSS methods, concluding that the wall similarity method is more suitable for fully rough flow \cite{bagherimiyab2013shear}. Thappeta (2023) applied nine classic methods in flume experiments to estimate bed shear stress and frictional velocity, concluding that the Saint-Venant approach provided the best stability, while turbulence models (TKE, TKE w', RSS) tended to underestimate the results \cite{thappeta2023bed}. Reynolds shear is often considered the best tool for evaluating bed shear stress \cite{nezu1994turbulence}. Biron (2004) and White (2022) used flow structures like baffles or geotextile bags to create complex flow fields, employing RSS as a calibration standard to optimize TKE and TKE w' \cite{biron2004comparing, white2022comparison}. However, assuming a positive correlation between scour intensity and bed shear stress, the TKE model, which captures the three-dimensional fluctuations of turbulent structures, yields predictions that better match the nonlinear coupling between scour intensity and shear stress compared to RSS. Recent research by Jeon and Kang (2024) used high-resolution LES data to evaluate the predictive capabilities of RSS, TKE, and the Logarithmic Law in complex flows. They found that RSS could reproduce LES-derived bed shear stress distributions and that the accuracy of RSS was affected by the inclusion of the transverse RSS component and the selection of extrapolation techniques \cite{jeon2024comparative}.

Current research on frictional flow velocity estimation models primarily centers on improvements to uniform flow models and innovations in numerical simulation techniques. However, the applicability of classical methods under complex flow conditions remains uncertain, with notable discrepancies in conclusions across different research contexts. Additionally, there is a lack of investigation into the complex flow patterns beneath riverbank sediment deposits. The flow field changes induced by both riverbank sediment deposits and flow diversion plates are comparable, reducing the effective flow cross-section and causing the flow around the structure to exhibit marked three-dimensional characteristics. Debris deposits from geological disasters often form a fan-shaped distribution. High outflow energy tends to increase the proportion of river width occupied by the deposits, posing a significant threat to upstream and downstream river sections during flood seasons \cite{zhou2018experimental}. This study constructs typical riverbank sediment deposits using a fixed-bed flume experiment and assesses the applicability of the classical frictional velocity model (Table 1) for estimating bed shear stress in affected river sections based on near-bed flow velocity data. The results contribute to sediment transport research and guide key parameters in post-earthquake river disaster simulations.

\begin{table}[H]
    \centering
    \small  % 使用小一号字体
    \caption{Methods for estimating shear velocity and related parameters}
    \label{tab:shear_stress_methods}
    \begin{tabular}{p{3.5cm}p{6cm}p{4cm}p{2.5cm}}
        \toprule
        \textbf{Method name} & \textbf{Equation} & \textbf{Comments} & \textbf{Citations} \\
        \midrule
        Reynolds shear stress (single-point) (RSS\_single) & 
        \( u_{*} = \sqrt{-\langle u'w' \rangle} \) &
        Assumes Reynolds stress is proportional to mean shear stress &
        \cite{graf1995bed} \\
        \midrule
        Reynolds shear stress (linear extrapolation) (RSS\_ext) &
        \( u_{*} = \sqrt{(-\langle u'w' \rangle)}_{z \to 0} \) &
        Assumes a linear distribution of Reynolds stress in the main flow area &
        \cite{graf1995bed} \\
        \midrule
        Turbulence Kinetic Energy (TKE) &
        \begin{minipage}[t]{6cm}
            \( \text{TKE} = \dfrac{\langle u'^2 \rangle + \langle v'^2 \rangle + \langle w'^2 \rangle}{2} \) \\[4pt]
            \( u_{*} = \sqrt{C_1 \cdot \text{TKE}}  \quad C_1 = 0.19 \)
        \end{minipage} &
        Assumes linear relationship between turbulent velocity fluctuations and mean shear velocity; assumes coefficient applied to rapidly varying flows &
        \cite{soulsby1983chapter} \\
        \midrule
        Vertical Turbulence Kinetic Energy (TKE\_w') &
        \( u_{*} = \sqrt{C_2 \langle w'^2 \rangle}  \quad C_2 = 0.9 \) &
        Same as for the TKE method &
        \cite{kim2000estimating} \\
        \midrule
        Logarithmic Law &
        \begin{minipage}[t]{6cm}
            \( u^+ = \frac{1}{\kappa} \ln(y^+) + A \) \\[4pt]
            \( u^+ = \frac{u}{u_{*}}  \quad y^+ = \frac{z}{z_0} \)
        \end{minipage} &
        Depth-averaged Law of the Wall; assumes developed boundary-layer velocity profile &
        \cite{schlichting1980boundary} \\
        \midrule
        Drag Shear Stress (DSS) &
        \( u_{*} = \sqrt{C_d U^2} \) &
        The quadratic stress law which relates the average shear stress at the bed to the square of the average fluid velocity &
        \cite{schlichting1980boundary} \\
        \midrule
        Global Resistance &
        \begin{minipage}[t]{6cm}
            When \( B/h_0 < 5.2 \) \\[4pt]
            \( u_{*} = \sqrt{g \delta J} \) \\[4pt]
            \( \dfrac{\delta}{h_0} = 0.44 + 0.106 \dfrac{R}{h_0} + 0.05 \sin\left[ \dfrac{2\pi}{5.2} \dfrac{B}{h_0} \right] \)
        \end{minipage} &
        Balance between the component force of the gravity component of the flow in the flow direction and the friction resistance of the bed surface &
        \cite{wang2001velocity} \\
        \bottomrule
    \end{tabular}
\end{table}

\section{Methods}
\subsection{Experimental Setup}
The experiment was conducted in the Sediment Testing Hall of the State Key Laboratory of Hydraulics and Mountain River Development at Sichuan University. The flume is 0.5 m wide, 0.42 m high, and 18 m long, with a smooth cement bottom and tempered glass sidewalls. The slope is 1‰ (Fig. 1(a)). Water is supplied by a circulating system, with flow controlled by valves. A stilling basin at the weir inlet dissipates energy, and a tailgate baffle at the outlet controls water depth (Fig. 1(c)). Fig. 1(d) shows 19 measurement sections (1\#-19\#), with 4-7 measuring lines (A-G) at each, and 7--10 vertical measurement points per line, depending on water depth.

\begin{figure}[H]
\centering
\includegraphics[width=0.45\textwidth]{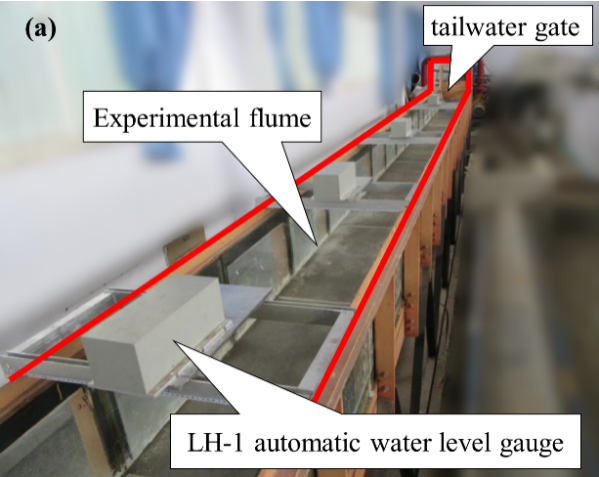}
\includegraphics[width=0.45\textwidth]{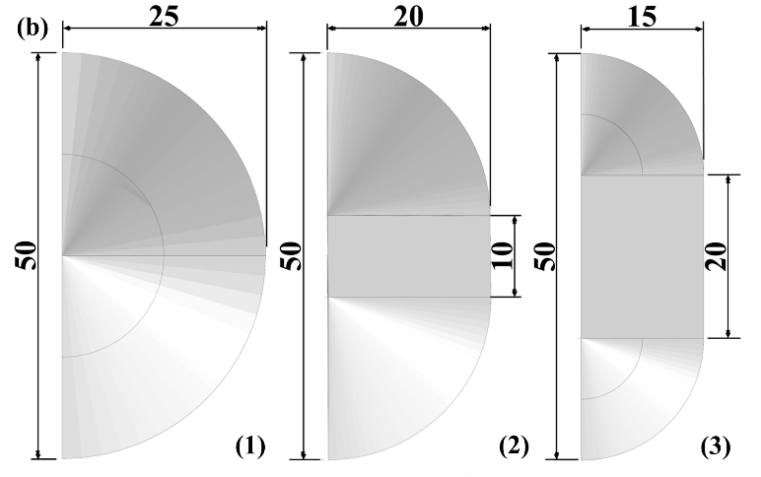}
\\
\includegraphics[width=0.9\textwidth]{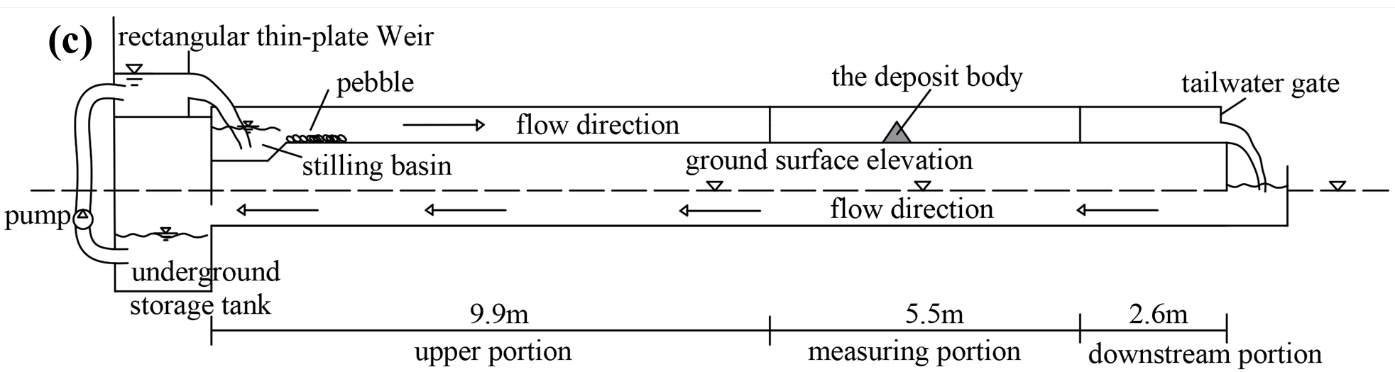}
\\
\includegraphics[width=0.9\textwidth]{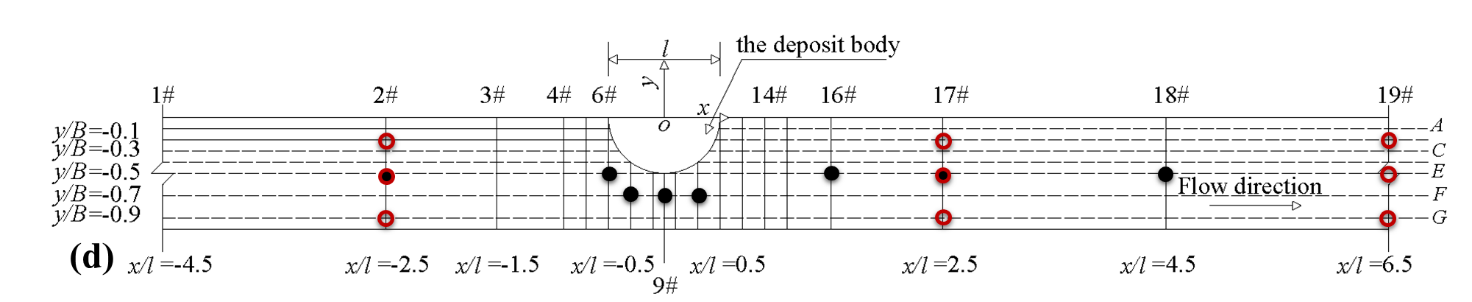}
\caption{(a) Photograph of the flume; (b) Three-dimensional schematic diagram of deposit body models((1)M1:$b/B$=0.5, $S$=45°; (2)M2:$b/B$=0.4, $S$=45°; (3)M3:$b/B$=0.3, $S$=45°); (c) Schematic illustration of the flume(The deposit in the figure is M1); (d)The plan view of the flume section with measurement points( The red hollow points indicate measurement locations under uniform flow conditions, while the black solid points correspond to measurement locations under deposit body conditions.)}
\end{figure}

This experiment is a generalized study based on field surveys conducted on mountainous rivers. The results show that under steep terrain constraints, the width-to-depth ratio of the cross-section is relatively small. The generalized deposit body model, summarized from the work of Li (2016) features lateral slopes between 30° and 45° and an occupied width ratio ($b/B$) ranging from 0.05 to 0.5 \cite{li2016study}. In this study, an acrylic model with a 45° slope was chosen, and the occupied width ratios ($b/B$) were set at 0.3, 0.4, and 0.5. The model length along the flow direction ($l$) was 50 cm, and the model's width ($B$) for the vertical walls was 15 cm, 20 cm, and 25 cm (Fig. 1b). After determining the deposit body dimensions and the width of the experimental flume, the flow rate range for the experiment was calculated based on the water depth and combined with the seasonal flow variations of mountainous rivers. The final flow rates used in the experiment were 31.9 L/s, 45 L/s, and 57.7 L/s. The origin of the coordinate system is at the intersection of section 9\# with the left bank and the flume bed. $x$ denotes the longitudinal distance parallel to the wall, and y denotes the transverse distance perpendicular to the wall. The dimensionless parameter $x/l$ indicates the relative position of the section to the accumulation body, and $y/B$ represents the relative position of the measurement point to the left bank. Uniform flow tests and corresponding accumulation body river tests were conducted. For uniform flow tests (series A1-A3), velocity measurements were taken at the red hollow points in Fig. 1(d). For accumulation body tests (series B1-B5), velocity measurements were taken at all 19 sections shown in Fig. 1(d). However, after verifying measurement consistency, only the solid black points in Fig. 1(d) were selected for frictional flow velocity estimation. Table 2 lists the experimental conditions.

\begin{table}[H]
\centering
\caption{Summary of Experimental Conditions}
\begin{tabular}{ccccc}
\toprule
Sequences & $Q$ (L/s) & The deposit body & $h_0$ (cm) & $B/h_0$ \\
\midrule
A1 & 31.9 & / & 11.6 & 4.31 \\
A2 & 45 & / & 14.75 & 3.39 \\
A3 & 57.7 & / & 17.6 & 2.84 \\
B1 & 31.9 & M1 & 11.6 & 4.31 \\
B2 & 45 & M1 & 14.75 & 3.39 \\
B3 & 57.7 & M1 & 17.6 & 2.84 \\
B4 & 31.9 & M2 & 11.6 & 4.31 \\
B5 & 31.9 & M3 & 11.6 & 4.31 \\
\bottomrule
\end{tabular}
\end{table}

\subsection{Experimental data processing}
This study utilizes the LH-1 automatic water level gauge (Fig. 1(a)) to measure and record water depth, with a measurement accuracy of 0.1mm, a range of 1000mm, a resolution of 0.05mm, and an absolute error of 0.1mm + range (m) / 8mm. Each measurement point is recorded every 0.5 seconds, with a total duration of 2 seconds per measurement, ensuring a complete capture of water surface fluctuations. Water depth analysis is based on time-averaged depth ($\Delta t$ = 2s).

Flow velocity is measured using a three-component Sontek ADV (Acoustic Doppler Velocimeter), with a sampling rate of 50 Hz. The device is equipped with both upward and downward probes. The downward probe has a blind spot within 5cm below the water surface, while the upward probe has a blind spot within 8cm above the bed surface. Consequently, when the water depth is $\geq$13cm, full water depth velocity measurements are possible. For depths $<$13cm, flow velocity beneath the water surface cannot be measured. According to Nortek (2018), the signal-to-noise ratio (SNR) and correlation coefficient (COR) have significant impacts on ADV flow velocity measurements: a low SNR indicates fewer particles or "scattering materials" in the water, leading to less reliable data, and when the COR is $<$70\%, the data lack strong correlation \cite{nortek2018comprehensive}. To ensure data reliability, measurements with an SNR $<$5dB and a COR $<$70\% are excluded.

\section{Results}
\subsection{Method Evaluation under Uniform Flow Conditions}
This paper initially estimates the near-bed friction velocity under the uniform flow conditions of several classical models (sequences A1-A3), based on experimental data, and presents a comparison of the results, as shown in Fig. 2. The variation in the experimental discharge is minimal, with the uniform flow Froude number ranging from 0.50 to 0.52. The results indicate that the estimates from the RSS\textsubscript{(single)} method are consistently lower than those from the RSS\textsubscript{(ext)} method, with deviations ranging from 7.1\% to 13.6\%. When compared to the DSS method, the estimation error remains within 5.4\%. This discrepancy arises because the DSS method's resistance coefficient is derived from the RSS\textsubscript{(single)} method. Among all the estimation models, the ratio of the highest estimate from the Global Resistance method to the lowest estimate from the TKE w' method increases with discharge, 1.2, 1.3, and 1.6 times, respectively. When excluding the Global Resistance method, the Logarithmic Law method consistently provides the highest estimates. This outcome aligns with the conclusions of previous uniform flow experiments \cite{biron2004comparing,jeon2024comparative}, thus confirming the reliability of the results obtained in this study.

\begin{figure}[H]
\centering
\includegraphics[width=0.8\textwidth]{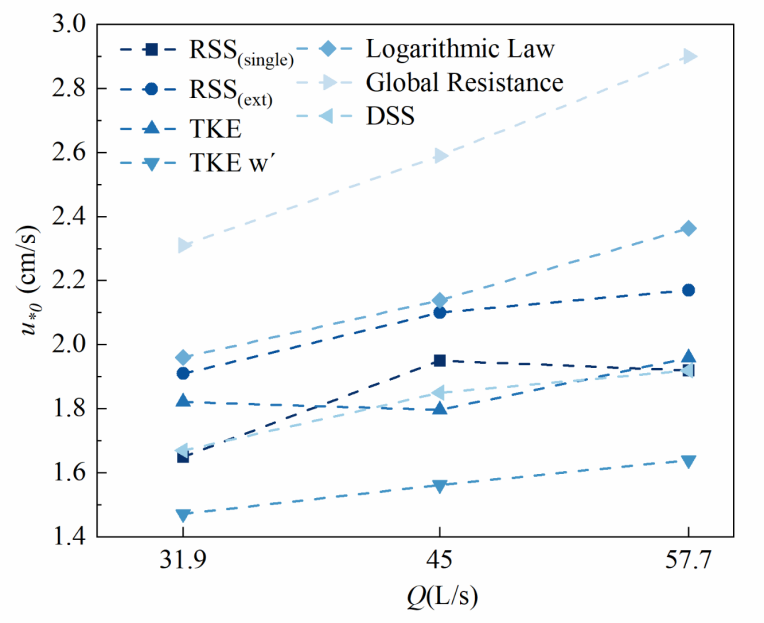}
\caption{Estimation of Near-Bed Friction Velocity under Uniform Flow Conditions}
\end{figure}

Qian and Wan (1999) synthesized the experimental findings of Nikitin, Grass, and Laufer on the turbulence characteristics of water flow, concluding that the vertical fluctuation intensity in the core flow region is nearly equal to the friction velocity (i.e., $\sqrt{\overline{w'^2}} / u_{*} \approx 1$) \cite{qian1999mechanics}. Similarly, Chen (2008) observed in flume tests that, under uniform flow conditions, the vertical turbulence intensity remains approximately constant with depth and is close to the friction velocity \cite{chen2008turbulence}. Based on the findings of these studies, this paper proposes using the ratio of the friction velocity to the measured vertical fluctuation intensity as a dimensionless parameter. The number of data points where this ratio lies between 0.85 and 1.15 for each model is then counted. $\sqrt{\overline{w'^2}} / u_{*0}$  larger count indicates a higher reliability of the model.

Fig. 3 presents the statistical results for the number of friction velocity values, estimated by different methods, that fall within the dimensionless range of 0.85 to 1.15. The comparison shows that the number of successful estimates using the TKE w' method and the Global Resistance method is considerably lower than that of the other methods, with fewer than half the number of successful estimates compared to the other methods. Among them, the Global Resistance method has the lowest count. The number of successful estimates using the RSS, TKE, Logarithmic Law, and DSS method are similar under various inflow conditions, with the percentage of successful estimates ranging from 54\% to 70\%. Notably, the DSS method and the RSS\textsubscript{(single)} method have the same statistical count, as the results of the former are based on the drag coefficient calculation from the latter. Based on the statistical analysis of the number of estimates for $\sqrt{\overline{w'^2}} / u_{*0}$ falling between 0.85 and 1.15, this study concludes that, under uniform flow conditions, the friction velocity estimates from the RSS method, TKE method, and Logarithmic Law method exhibit higher accuracy. These findings align with the research of several scholars \cite{biron2004comparing,white2022comparison,jeon2024comparative}.

\begin{figure}[H]
\centering
\includegraphics[width=0.8\textwidth]{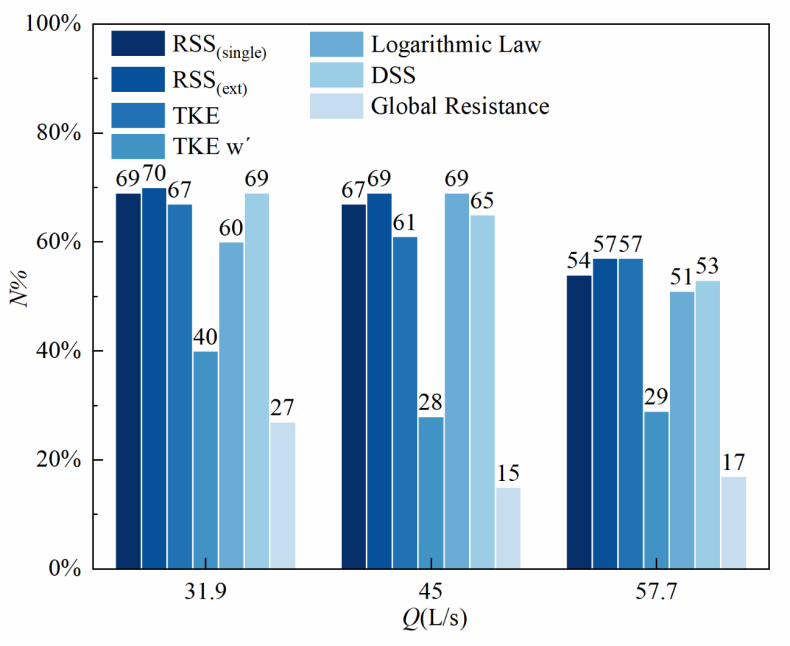}
\caption{Percentage Distribution of Data Points (\%) for $\displaystyle \frac{u_*}{\sqrt{\overline{w'^2}}} = 0.85 \sim 1.15$}
\end{figure}

\subsection{Method Evaluation under Deposit body Conditions}
The presence of a deposit body in a straight river channel causes significant changes in the surrounding flow field. Upstream, far from the deposit body, the flow velocity decreases, and the water level rises. Upon entering the river section with the deposit body, the flow direction gradually shifts as the shape of the deposit body evolves. The cross-sectional area for flow decreases, causing an increase in flow velocity and the development of sinking flow, resulting in a flow field with prominent three-dimensional characteristics.

Downstream of the deposit body, the flow cross-section gradually recovers, and flow dispersion occurs. A backflow region is formed downstream near the deposit body. Influenced by the high-velocity mainstream region, water near the deposit body is carried away, while low-velocity flow near the bank wall moves upstream to feed the backflow region. The flow becomes turbulent, with the formation of numerous bubbles and vortices. The flow near the deposit body can be categorized based on the flow connection patterns and velocity characteristics as follows: Back water region, Stagnant region, Mainstream region of the deposit body, Mainstream region downstream of the deposit body, backflow region, and Downstream recovery region. A schematic diagram of the flow divisions is shown in Fig. 4.

\begin{figure}[H]
\centering
\includegraphics[width=0.9\textwidth]{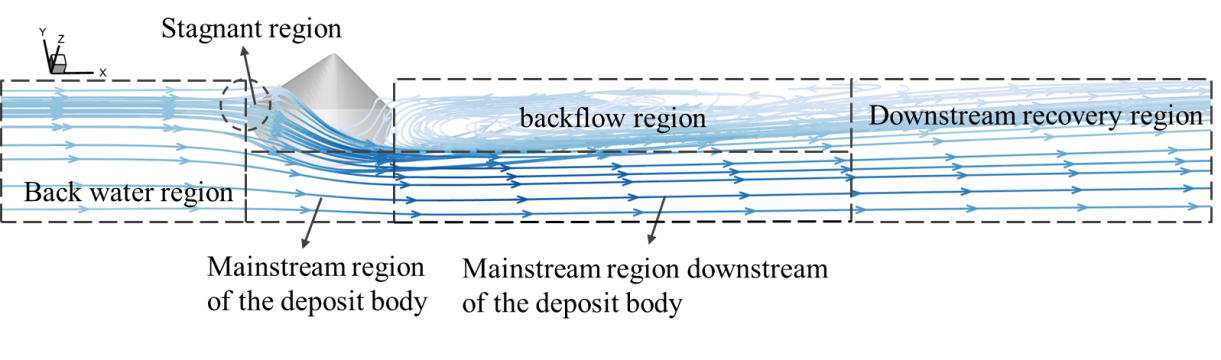}
\caption{Presents a schematic representation of the river flow zonation within the deposit body.}
\end{figure}

The presence of the deposit body influences both the flow regime and the flow structure across different regions. In comparison with flow characteristics under uniform flow conditions, the deposit body induces changes in frictional behavior. In mountainous rivers, bedload movement is strongly linked to the drag force near the riverbed. Therefore, analyzing the frictional properties of the flow field near the deposit body is essential for understanding its effect on sediment transport.

Experimental analysis under uniform flow conditions reveals that, among the four methods, the Global Resistance method exhibits the lowest reliability, while the primary drawback of the Global Resistance method is its inability to independently determine the resistance coefficient. Given that the river channel containing the deposit body is under non-uniform flow conditions, the Reynolds stress distribution along the vertical line no longer adheres to a linear pattern \cite{song1994non}. As a result, the RSS\textsubscript{(ext)} method is unsuitable for estimating the frictional velocity in the flow field near the deposit body. Li (2016) demonstrated that the flow in both the upstream and deposit body sections follows a logarithmic distribution with respect to water depth. However, the distribution parameters are affected by flow rate and deposit body morphology \cite{li2016study}. This suggests that the Logarithmic Law method is still applicable in river channels containing deposit bodies. Consequently, in this study, velocity data within $y/h \leq 0.2$ were used in the distribution formula for calculation.

Based on the previous analysis, four methods are employed to estimate the frictional velocity in the flow field near the deposit body: the RSS\textsubscript{(single)}, TKE, TKE w', and Logarithmic Law method. When $Q$ = 31.9--45 L/s, certain sections of the deposit body exhibit limited effective velocity data due to shallow water depths. Data analysis and comparison indicate that the velocity distribution pattern under different flow conditions is consistent. To maintain data completeness, this study uses the B3 condition ($Q$ = 57.7 L/s, $b/B$ = 0.5, $S$ = 45°) for the analysis of friction distribution characteristics in the river section near the deposit body ($x/l$ = -2.5 to 4.5), as illustrated in Fig. 5-6.

Fig. 5 illustrates the horizontal flow vectors and longitudinal velocity contour plots at $z$ = 0.5 cm in the river channel with the deposit body. The maximum longitudinal velocity occurs in the mainstream region downstream of the deposit body, specifically around $y/B$ = -0.5 and $x/l$ = 0.9. As shown in Fig. 6, the frictional velocity in the backwater region is consistently lower than in other areas due to the reduced longitudinal velocity caused by the backwater effect. However, variations in the distribution of results from different methods are observed in the deposit body section and downstream of the deposit body.

Fig. 6(a) and (d) present the frictional velocity contours obtained using the RSS\textsubscript{(single)}, TKE, TKE w', and Logarithmic Law method. Comparison shows that the peak values for both the TKE and TKE w' methods occur on the left side downstream, specifically at $y/B$ = -0.4, $x/l$ = 1.1, and $y/B$ = -0.1, $x/l$ = 2.4, respectively. The frictional velocity decreases from left to right across the cross-section. The peak values for the RSS\textsubscript{(single)} and Logarithmic Law method are found on the right side downstream, at $y/B$ = -0.7, $x/l$ = 1.6, and $y/B$ = -1, $x/l$ = 2.6, respectively. In the case of the former, the frictional velocity decreases from the peak value towards both banks across the cross-section, while in the latter, it decreases towards the left bank.

Assuming that regions of high flow velocity correspond to areas with high friction velocity, the distribution of friction velocity is expected to exhibit a certain degree of similarity to the flow velocity distribution. A comparison of Fig.5 and 6 reveals that the friction velocity estimated using the TKE method aligns closely with the contour map of longitudinal velocity, showing strong consistency in both peak values and trends.

\begin{figure}[H]
\centering
\includegraphics[width=0.8\textwidth]{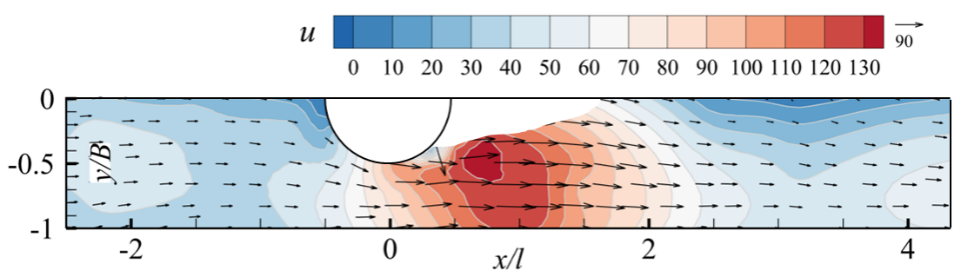}
\caption{Velocity vectors and longitudinal velocity contour map for experimental group B3 (cm/s)}
\end{figure}

\begin{figure}[H]
\centering
\includegraphics[width=0.9\textwidth]{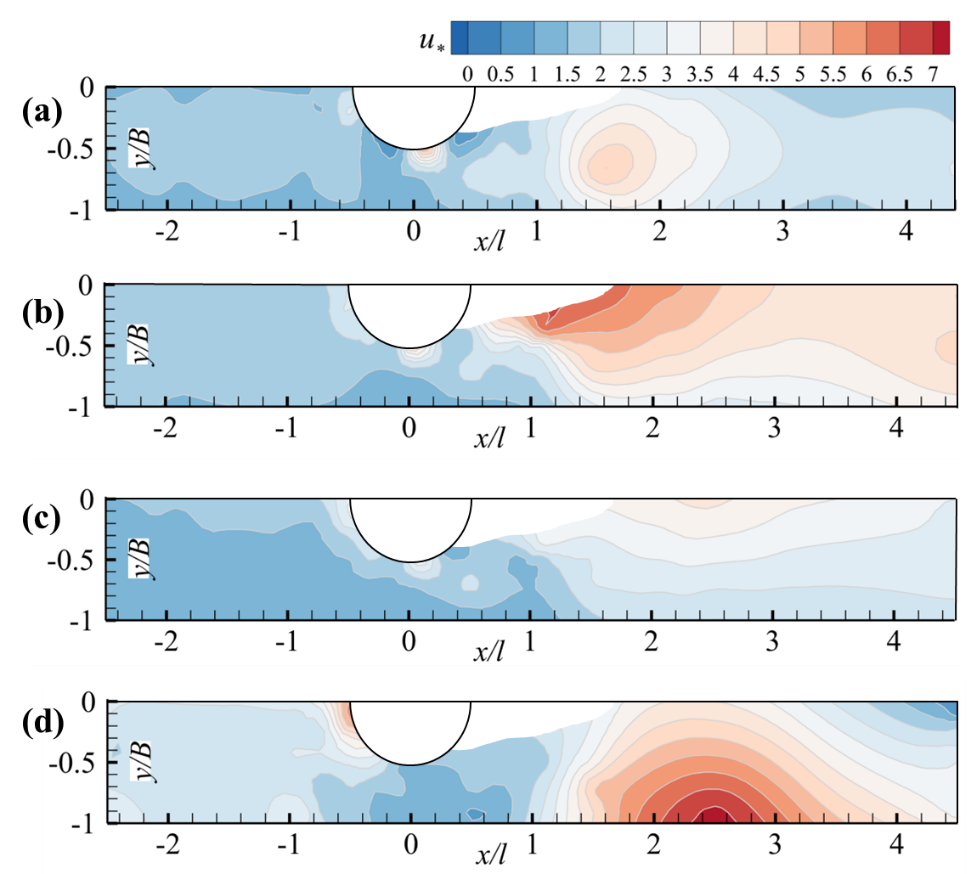}

\caption{Contour map showing the distribution of friction velocity for experimental group B3 (cm/s); (a) RSS\textsubscript{(single)} method; (b) TKE method; (c) TKE w' method; (d) Logarithmic Law method}
\end{figure}

\subsection{Analysis of Frictional Characteristics in Deposit Body Channels}
The deposit body influences both the adjacent flow field and the bed shear stress distribution near the bed surface, which significantly affects the bed evolution of mountain rivers dominated by bedload transport. To assess the changes in bed friction characteristics following the formation of the deposit body, this study evaluates the ratio of the friction velocity $u_*'$, derived from the model, to the friction velocity $u_{*0}$ of uniform flow with the same discharge, using it as a dimensionless indicator of friction characteristics. The analysis will further examine the variation in friction characteristics across different model estimates. Based on the influence of the deposit body on the river flow field and its flow regime zoning characteristics, several measurement lines are selected to represent: 1 Back water region, 2 Mainstream region of the deposit body, 3 Mainstream region downstream of the deposit body, and 4 Downstream recovery region. The near-bed friction velocity for these representative measurement lines is estimated using four classic models. The study further compares and analyzes the relationship between friction characteristics, flow intensity, and the width of the deposit body.

The representative measurement lines for the back water region are 2-E and 6-E. The flow velocity in this area is lower than that under uniform flow conditions, and the flow regime is relatively stable, with negligible longitudinal and transverse slopes \cite{huang2017experimental}. Fig.7 and 8 indicate that the friction velocity in the deposit body section, as estimated by each model, is approximately 0.75 to 1.25 times that under uniform flow conditions. Additionally, factors such as flow intensity and the width of the deposit body have little impact on the friction velocity. Compared to other river sections, the friction velocity variation in this area is small, suggesting that in the back water region, where turbulence intensity is low, the flow regime is close to that of uniform flow. The results from all models show minimal differences and demonstrate good applicability.

The main survey lines in the mainstream region of the deposit body are 7-F, 9-F, and 11-F, with the flow cross-section at 9-F being most affected by the constriction effect. Except for the Logarithmic Law method, the friction velocity estimates from the other three models show minimal variation compared to the backwater region before the flow cross-section begins to recover. A significant increase in friction velocity is only observed once the cross-section starts to recover. Interestingly, after recovery, the dimensionless friction velocity estimated using the logarithmic velocity method decreases. Under experimental conditions with an occupation width of M1, the dimensionless friction velocity is around 0.5. In this region, the influence of factors such as flow intensity and the occupation width of the deposit body on friction velocity remains unclear.

The representative survey lines in the mainstream region downstream of the deposit body are 16-E and 17-E. In this area, the flow cross-section downstream of the deposit body has fully recovered. High-velocity flow in the mainstream region and low-velocity flow in the backflow region intertwine, creating a shear zone that results in strong turbulence in the flow field. In this region, the friction velocity increases significantly. Depending on the experimental conditions and estimation methods, the peak friction velocity is approximately 1 to 4 times that under uniform flow conditions. It is evident that the flow intensity and occupation width of the deposit body influence the friction characteristics in a consistent manner: as flow intensity decreases and the occupation width of the deposit body increases, the near-bed dimensionless friction value at survey line 17-E becomes higher. As the height of the natural deposit body increases, its cross-section decreases, which results in a reduction in the proportion of flow blocked by the deposit body as the flow rate increases and the water level rises. Conversely, when the flow rate is low, the resistance of the deposit body becomes more pronounced, significantly affecting the flow field. At the same flow rate, a larger occupation width of the deposit body leads to a higher proportion of flow resistance in the river, amplifying the impact of the deposit body's resistance on the flow field.

The representative survey lines in the downstream recovery region are 18-E and 19-E. In this area, the backflow region with low-velocity zones no longer exists. The lateral velocity gradient of the river cross-section is small, and the turbulence intensity gradually weakens along the flow direction. The friction velocity returns to a range of 1 to 2.5 times that of uniform flow conditions. Fig. 7 illustrates that under low flow conditions, although the deposit body segment causes considerable flow resistance, the flow field in the downstream region recovers more rapidly to a uniform flow state. Fig. 8 shows that the occupation width of the deposit body does not exhibit a clear pattern of influence on friction velocity. This indicates that in the downstream recovery region, the impact of the deposit body gradually weakens, and the flow field approaches uniformity, with the effect of occupation width on friction velocity becoming negligible.

Overall, the near-bed friction velocity upstream of the deposit body is slightly lower than that under uniform flow conditions, until reaching the maximum constricted cross-section of the deposit body segment, where the dimensionless friction velocity is about 1 with little variation. As the river cross-section recovers, the friction velocity increases significantly, peaking in the mainstream region downstream of the deposit body, where it can reach 3 to 4 times the value under uniform flow conditions. In the downstream recovery region, the friction velocity gradually decreases, approaching a uniform flow state. The estimation results using the RSS\textsubscript{(single)}, TKE, and TKE w' method all accurately capture the changes in the bed surface friction characteristics of the deposit body river channel along the flow direction. However, the friction velocity estimated by the Logarithmic Law method in the mainstream region of the deposit body is lower than that in the backwater region, which contradicts the longitudinal distribution pattern of the flow velocity. From the perspective of experimental variables, the weaker the flow intensity and the larger the occupation width of the deposit body, the more pronounced the actual water resistance effect of the deposit body and the greater the disturbance to the original flow field. As a result, the increase in friction velocity in the mainstream region downstream of the deposit body will also be more significant.

\begin{figure}[H]
\centering
\includegraphics[width=0.9\textwidth]{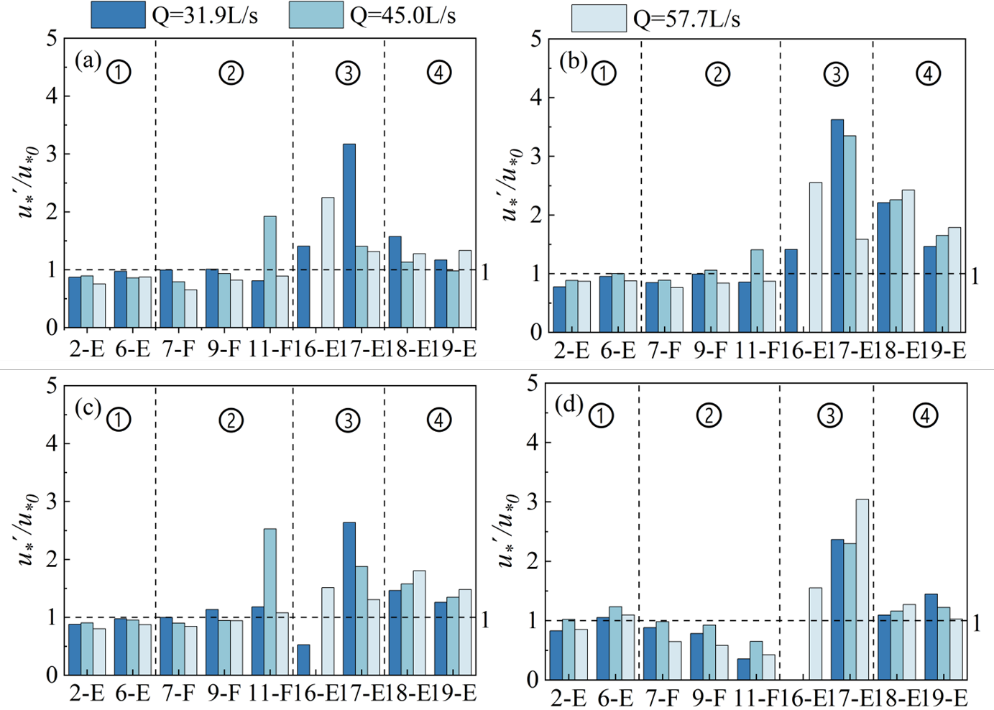}
\caption{Presents a comparison of results under varying inflow conditions (M1: $b/B$ = 0.5, $S$ = 45°)}
\end{figure}

\begin{figure}[H]
\centering
\includegraphics[width=0.9\textwidth]{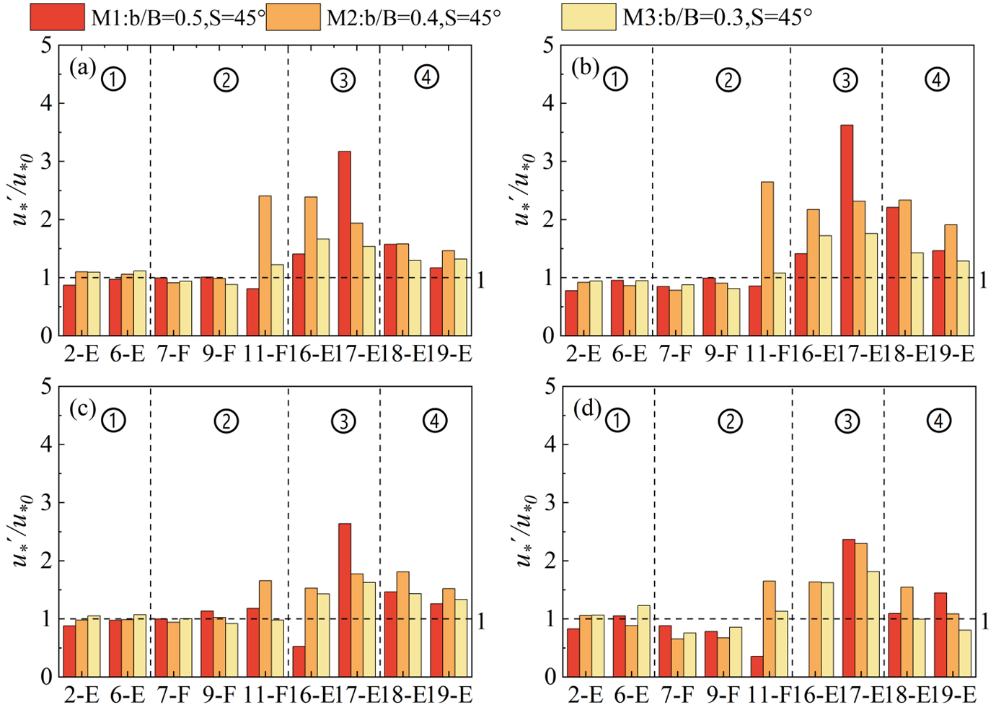}

\caption{Presents a comparison of results under varying intake ratio conditions ($Q$ = 31.9 L/s). (a) RSS\textsubscript{(single)} method; (b) TKE method; (c) TKE w' method; (d) Logarithmic Law method;1 Back water region; 2 Mainstream region of the deposit body; 3 Mainstream region downstream of the deposit body and 4 Downstream recovery region.}
\end{figure}

\section{Discussion}
The local shear stress estimation methods used in this study are based on measured flow velocities, so it is important to discuss the reliability of the ADV data. Previous studies \cite{mclelland2000new} have examined the impact of ADV instrument noise on turbulence measurements. Among these, Doppler noise has almost no effect on TKE, TKE w', and RSS, meaning Doppler noise does not influence the estimation results of TKE, TKE w', and RSS. As for sampling noise, the sampling frequency in this study is 50 Hz, which minimizes the sampling noise and reduces the risk of data aliasing, thereby improving the accuracy of the data. The missing data in some groups of the 16E measurement line can be attributed to the following two factors: In the B1 test group ($Q$ = 31.9 L/s, $b/B$ = 0.5, $S$ = 45°), the large velocity gradient near the bed surface ($y/h \leq 0.2$) led to a low correlation coefficient when fitting the data using the logarithmic velocity method, so the data was excluded. In the B2 test group ($Q$ = 45 L/s, $b/B$ = 0.5, $S$ = 45°), strong turbulence in the left recirculation zone generated bubbles, preventing the ADV from obtaining valid data under the experimental conditions.

To achieve a more continuous and smooth visualization, the friction velocity distribution contour map in Section 3.2 was generated using Tecplot software, utilizing the near-bed friction velocity estimates from all measurement lines and the Kriging interpolation method. In Section 3.3, the longitudinal variation of friction velocity was represented using only a selection of representative measurement lines for a quantitative comparison of different models and experimental groups. Therefore, due to the errors introduced during the interpolation process, there may be some discrepancies between the two sets of friction velocity results, particularly in regions where the data changes rapidly.

Under the influence of the deposit body, the variation in friction velocity is not only related to the near-bed flow velocity but also closely connected to the turbulence intensity of the flow field. In the deposit body section, the reduction in flow cross-section causes an increase in flow velocity. However, regardless of changes in flow intensity and occupation width, the friction velocity remains comparable to or even smaller than that of uniform flow, with no clear relationship observed between these two variables. The effects of flow intensity and occupation width on the friction characteristics are ultimately reflected in the downstream section of the river. Specifically, the higher the mainstream velocity and the greater the velocity gradient with the backflow region, the more intense the turbulence, which results in a near-bed friction velocity downstream of the deposit body that significantly exceeds that of uniform flow. These phenomena suggest that the bed friction characteristics influenced by the deposit body are mainly determined by turbulence intensity.

Based on the above analysis, the four classic evaluation models demonstrate good adaptability and minimal differences when applied to the analysis of friction characteristics in non-uniform flow fields of rivers with similar deposit bodies, particularly in back water region with low turbulence and in the narrowed river sections downstream of the deposit bodies. In the downstream recovery region, where the flow begins to recover and spread, the turbulence in the flow field intensifies, causing the vertical velocity distribution to deviate from the logarithmic profile. The logarithmic velocity method is unable to accurately represent the changes in friction characteristics. However, the RSS\textsubscript{(single)}, TKE, and TKE w' method can all capture the turbulence information of the flow, allowing them to reflect the impact of the deposit body on the friction characteristics. Among these, the TKE method provides a more comprehensive consideration of the turbulence characteristics.

In this study, both the flume and the model are based on generalized dimensions. The determination of the experimental flow rate takes into account two factors: (1) The test width-to-depth ratio is designed to approximate that of natural rivers, while also considering the seasonal flow characteristics typical of mountainous rivers; (2) Given the depth limitations of the flow velocity measurement equipment, the range of flow variations is relatively small. As a result, the flow intensity in the deposit body river section does not show a clear trend in friction velocity, except in the downstream recovery area.

\section{Conclusions}
This study performs a series of flume model experiments to evaluate the applicability of classical friction velocity estimation models in determining near-bed friction velocity in deposit body river sections. It also analyzes the changes in bed friction characteristics influenced by the deposit body and explores the underlying causes. The conclusions are as follows:

(1) Under uniform flow conditions, using vertical fluctuating intensity close to the friction velocity as the criterion, the classical friction velocity estimation models were quantitatively evaluated. It was concluded that the RSS, TKE, and Logarithmic Law methods provide relatively accurate estimates of friction velocity.

(2) The river section influenced by the deposit body is classified Back water region, Stagnant region, Mainstream region of the deposit body, Mainstream region downstream of the deposit body, Backflow region, and Downstream recovery region. Through the comparison and analysis of the contour maps of near-bed longitudinal velocity and friction velocity, it was observed that the friction velocity estimated using the TKE method shows good agreement with the contour map of longitudinal velocity, both in terms of peak values and overall trend.

(3) Through the analysis of the variation in the dimensionless friction indicator along the river section, the following conclusions are drawn: In non-uniform flow in deposit body river sections, the friction velocity is mainly influenced by the turbulence intensity of the flow, rather than the longitudinal time-averaged velocity. Compared to uniform flow conditions, the friction characteristics show little variation in the back water region of the deposit body and in areas with narrowed cross-sections. After the water flow cross-section recovers, the friction characteristics significantly increase due to intense turbulence caused by the flow expansion. Modeling approaches based on flow field fluctuation information, such as the RSS\textsubscript{(single)}, TKE, and TKE w' method, can effectively reflect the friction characteristics of non-uniform flow fields. When the flow intensity is low or the deposit body occupies a larger width, the higher actual flow resistance of the deposit body will lead to a greater increase in friction velocity in the downstream recovery region.

\section{Data availability}
The datasets generated during and/or analysed during the current study are available from the corresponding author on reasonable request.

\section{Nomenclature}
\begin{table}[H]
\centering
\begin{tabular}{p{2cm} p{10cm}}
$B$ & width of flume [m] \\
$b$ & bottom width of deposit [m] \\
$C_d$ & drag coefficient \\
$g$ & acceleration due to gravity [m/s$^2$] \\
$h$ & backwater depth [m] \\
$h_0$ & uniform water depth [m] \\
$L$ & length of deposit body along flow direction [m] \\
$Q$ & flow rate [L/s] \\
$S$ & lateral slope of deposit body [°] \\
$R$ & hydraulic radius [cm] \\
$U$ & velocity [cm/s] \\
$u_*$ & Friction velocity [cm/s] \\
$u_{*0}$ & Friction velocity of uniform flow [cm/s] \\
$u_0$ & uniform flow [cm/s] \\
$u_*'$ & Friction velocity of deposit body [cm/s] \\
$u'$ & velocity fluctuation of streamwise component [cm/s] \\
$v'$ & velocity fluctuation of lateral component [cm/s] \\
$w'$ & velocity fluctuation of vertical component [cm/s] \\
$\kappa$ & von Karman's constant \\
$A$ & constant of integration \\
$z$ & distance from bed surface [cm] \\
$z_0$ & characteristic roughness length [cm] \\
$J$ & hydraulic gradient \\
$\langle \rangle$ & denotes time averaging \\
$\rho$ & water density [kg/m$^3$] \\
$\tau$ & near-bed shear stress [pa] \\
\end{tabular}
\end{table}

\bibliography{main}

\section*{Acknowledgements}
The completion of this article was inseparable from the contributions of all authors. Their support is gratefully acknowledged. The authors would like to thank all the reviewers who participated in the review.

\section*{Author Contributions}
Z.J. was involved in methodology, data curation, funding acquisition, writing-original draft; Q.R.H. participated in Investigation, methodology and Writing-review and editing; G.Z.X. was responsible for methodology and writing---review and editing; X.H.T. took part in Investigation and writing-review and editing; Z.J.H. was responsible for methodology and writing-review and editing; B.Y. was responsible for methodology and writing-review and editing.

\section*{Funding}
This work was supported by the Natural Fund Project of Sichuan Science and Technology Department (Grant No. 2023NSFSC0950), National Natural Science Foundation of China (Grant No. 51409224), the Open Fund Research of State Key Laboratory of Hydraulics and Mountain River Engineering, China (Grant No. SKHL2217) and the Open Research Subject of Key Laboratory of Fluid Machinery and Engineering (Xihua University), Sichuan Province (Grant No. LTJX-2022002).

\section*{Declarations}
\subsection*{Competing interests}
The authors declare no competing interests.

\end{document}